# Robust, high brightness, degenerate entangled photon source at room temperature

M. V. Jabir, G. K. Samanta

*Photonic Sciences Lab., Physical Research Laboratory, Navarangpura, Ahmedabad 380009, Gujarat, India*
*Corresponding author: jabir@prl.res.in*

**We report on a compact, simple and robust high brightness entangled photon source at room temperature. Based on a 30 mm long periodically poled potassium titanyl phosphate (PPKTP), the source produces non-collinear, type-0 phase-matched, degenerate photons at 810 nm with pair production rate as high 39.13±0.04 MHz/mW at room temperature. To the best of our knowledge, this is the highest photon pair rate generated using bulk crystals pump with continuous-wave laser. Combined with the inherently stable polarization Sagnac interferometer, the source produces entangled state violating the Bell's inequality by nearly 10 standard deviations and a Bell state fidelity of 0.96. The compact footprint, simple and robust experimental design and room temperature operation, make our source ideal for various quantum communication experiments including long distance free space and satellite communications.**

Entangled photon sources, basic ingredient for many quantum optical experiments, are of paramount importance not only for the fundamental research [1] but also for a variety of applications in real world quantum communications [2] and quantum computing [3]. However, the realization of next generation envisaged projects towards the implementation of world-wide quantum network through the ground to satellite and or inter-satellite links [4, 5]require development of compact and robust entangled photon sources with high brightness and entanglement visibility. Over decades, a variety of schemes have been proposed and implemented [6, 7] for entangled photons, however, the polarization entangled photon sources realized through the spontaneous parametric down-conversion (SPDC) in second order, ($\chi^2$),bulk nonlinear crystals remains the most appropriate choice.

Since, the parametric gain of $\chi^2$ nonlinear processes are low, efforts have been made to improve the brightness of the entangled photon sources by using different nonlinear crystals in bulk [8, 9]and waveguide[10]structures, in different phase-matching geometries including type-I, type-II and type-0, and different experimental schemes. As, the length and nonlinearity are two important crystal parameters highly influencing the gain of the SPDC process for a given pump laser, the brightness of the entangled photon sources have been significantly improved over the years through the use of long periodically poled crystals engineered for high effective nonlinearity.

As such, use of type-0 phase matching of periodically poled potassium titanyl phosphate (PPTKP) crystal in crossed-crystal geometry [11] have produced collinear, on-degenerate photon pairs at a rate as high as 0.64 MHz/mW. While one can expect further enhancement in the pair photon rate by operating the source at degeneracy, however, the separation and detection of individual photons of the collinear, co-polarized photon pairs at degenerate wavelength is the major challenge to overcome. Attempt has been made to overcome such problem with the use of non-collinear, type-0, third order quasi-phase-matching (QPM) in periodically poled lithium tantalite (PPSLT) crystal[12], however, use of third order QPM resulted in moderately low rate of photon pairs (98.5 kHz/mW). On the other hand, the requirement of active temperature control for non-critical phase-matching in periodically poled crystals increases the overall complexity of the sources. However, the success of future quantum communication experiments, in addition to other obstacles, demands engineering of high brightness entangled photon sources with minimal system complexities. Here, we demonstrate high brightness entangled photon source producing on-collinear, degenerate photon at 810 nm with a photon pairs rate as high as 39.13±0.04MHz/mW at room temperature. Based on a single, 30 mm long PPKTP crystal configured in a polarization Sagnac interferometer, the source produces entangled state violating the Bell's inequality by nearly 10 standard deviations and a Bell state fidelity of 0.96.

The schematic of the experimental setup is shown in Fig. 1. A continuous-wave, single-frequency (<12 MHz) UV laser providing 100 mW of output power at 405 nm is used as a pump laser. The laser power to the experiment is controlled using a half-wave plate ($\lambda/2$) and a polarizing beam splitter (PBS) cube. Use of single mode fibre (SMF) improves the spatial beam quality of the pump. A convex lens of focal length, $f$=300 mm, is used to focus the pump beam at the center of the crystal to waist radius of ~145 μm. A 30 mm long, 2×1 mm$^2$ in aperture, single grating, PPKTP (Raicol, Israel) crystal of period, $\Lambda$=3.425 μm, is used for type-0 (e→e+e) phase-matched down conversion of pump beam at 405 nm. To generate entangled photons, we have devised a novel experimental scheme based on polarization Sagnac interferometer comprising with a dual-PBS, dual-$\lambda/2$ plate designed for both 405 nm and 810 nm, and two high reflecting (R>99% at 405 nm and 810 nm) mirrors, M. The working principle of the scheme can be understood as follows. The $\lambda/2$ plate placed before the dual-PBS controls the polarization of the pump beam in such a way that the reflection (vertical polarization, V) and transmission (horizontal polarization, H) ports of the dual-PBS have equal laser powers. The V polarized pump photons travelling in clock-wise (CW) direction in the Sagnac interferometer, generate V polarized SPDC photons in the PPKTP crystal owing to type-0 phase-matching condition. The dual-$\lambda/2$ transforms the polarization of both the pump and SPDC photons from H to V and vice versa. Therefore, the V polarized SPDC photons propagating in CW direction of the Sagnac interferometer transformed into H polarized photons and pass through the dual-PBS. At the same time, the H polarized pump photons travelling in the counter clockwise direction (CCW) in the Sagnac interferometer transformed into V polarization after passing through the dual- $\lambda/2$ plate and generate V polarized SPDC photons in the same PPKTP crystal. The dual-PBS combines the SPDC photons generated in both CW and CCW directions of Sagnac interferometer. Since both the CW and CCW pump beams follow the same path in opposite directions, and the PPKTP crystal is placed at a position symmetric to the dual-PBS, the present experimental scheme is robust against any optical path changes to produce SPDC photons in orthogonal polarizations with ultra-stable phase. Using apertures, high reflecting mirrors, and interference filter (IF) having transmission bandwidth of ~10 nm centered at 810 nm, the SPDC photons are collected and subsequently detected with the help of single photon counting module (SPCM-AQRH-14-FC Excelitas). Analyzer, A, comprising with a PBS and a half-wave plate, is used to analyze the polarization entanglement of the generated photons.

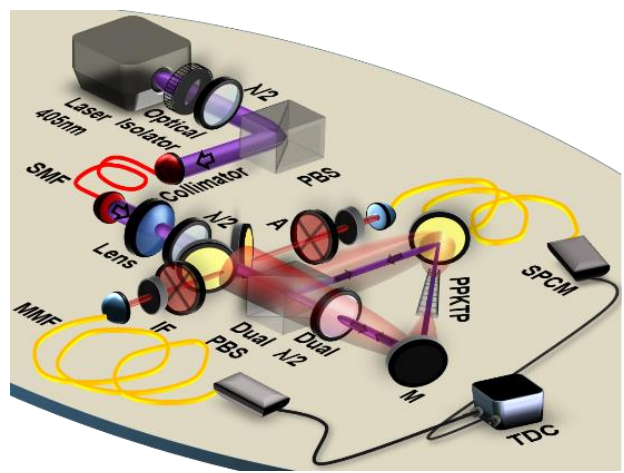

*Fig.1. Schematic of the experimental setup. λ/2, half-wave plate; PBS, polarizing beam splitter cube; Dual PBS, dual λ/2: polarizing beam splitter cube and half-wave plate at 405 nm and 810 nm; M, Mirrors; PPKTP, nonlinear crystal; A, analyzer; IF, interference filter; SPCM, single photon counting module; TDC, time-to-digital converter.*

To verify the generation of degenerate SPDC photons in non-collinear, type-0 phase matching, we pumped the crystal at an input power of 40 mW. Using a CCD camera (SP620U, Spiricon) along with an interference filter of spectral width ~10 nm centered at 810 nm, we have recorded the angular spectrum or transverse momentum distribution of the SPDC photons using a convex lens of focal length, $f$ = 50 mm in $f$-$f$ optical system configuration[13]. Since each pixel represents a particular transverse momentum of the SPDC photons, using the recorded spatial distribution of the SPDC photons we calculated the emission angle (half cone angle) of the photons at different crystal temperature with the results shown in Fig. 2. As evident from Fig. 2, the emission angle of the SPDC photons decreases from 2.15º to 0º with the increase of crystal temperature from 22.2ºC to 35.5ºC, clearly verifying the transition of the phase-matching of the SPDC photons from non-collinear to collinear geometry. While decrease in crystal temperature below 22ºC results in further increase in the non-collinear phase-matching angle of the degenerate SPDC photons, the increase of crystal temperature beyond 35º results non-degenerate SPDC photons in collinear phase-matching geometry. A linear fit to the experiment data for the crystal temperature variation across 22$^0$C to 29ºC reveals that the emission angle (half cone angle) of the SPDC photons change by 0.037º for a change of 1ºC in the crystal temperature. Such small variation in the emission angle with crystal temperature indicates the possibility of generation of entangled photons at room temperature without using any active temperature controls commonly required for non-critical phase-matching[14]. Using the Sellmeier equations [15] of PPKTP crystals and the equations for non-collinear phase-matching we theoretically calculated (solid line) the emission angle of the SPDC photons in good agreement with the experimental data (dots). The non-collinear generation and variation of the SPDC ring with crystal temperature is also evident from the CCD images of the intensity distribution of the SPDC photons as shown by the inset of Fig. 2.

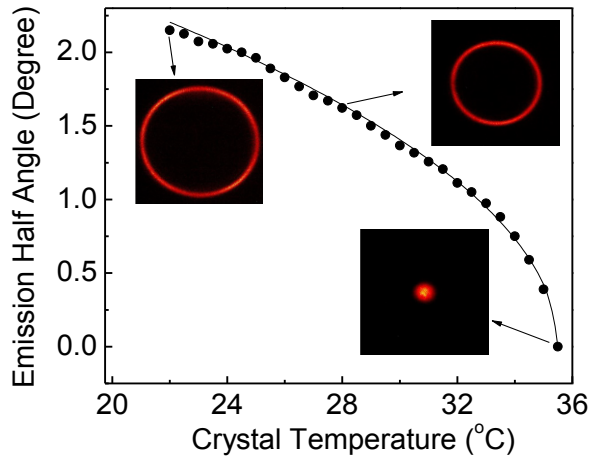

*Fig. 2. Variation of emission half-angle of SPDC photons measured in free space with crystal temperature. Solid line corresponds to the theoretical fit to the experimental data. (Inset)CCD images of the SPDC ring at three crystal temperatures, 22ºC, 28ºC, and 35ºC.*

We also studied the spectral distribution of the SPDC source at both non-collinear and collinear phase-matching geometries. Pumping the crystal at pump power of 40 mW and removing the interference filter we have measured the spectrum of the SPDC photons using a spectrometer (HR 4000, Ocean Optics) while changing the crystal temperature. The results are shown in Fig. 3. As evident from Fig. 3, the increase in crystal temperature from 29ºC results in non-collinear, degenerate SPDC photons at 810 nm up to the temperature of 35ºC, however, further increase in crystal temperature up to 65ºC produces collinear, non-degenerate SPDC photons with wavelength tunability over 810-743 nm in the signal wavelength (open circles) and 810-891 nm in the idler (solid circle). The solid line is the tuning curve predicted from the Sellmeier equations of PPKTP crystal [15], confirming a reasonable agreement between experimental data and theoretical calculations. The inset Fig. 3 shows the measured spectra of collinear SPDC photon pairs at crystal temperature of 35º, 40º, 45º, 50º, 55º, and 60ºC. It is evident from the inset of Fig. 3, the source produces degenerate SPDC photons at a spectral width (full width at half maximum, FWHM) of ~31 nm centered at 810 nm, with a signature of high parametric gain of the PPKTP crystal. We also observed similar spectrum at crystal temperature below 35ºC. However, the spectral width of the signal and idler photons decrease with the increase of crystal temperature away from degeneracy. From the normalized intensity of the SPDC photons, it is also evident that the development of SPDC source with high brightness requires operation of the periodically poled crystal based sources in degenerate, non-collinear, type-0 phase matching geometry.

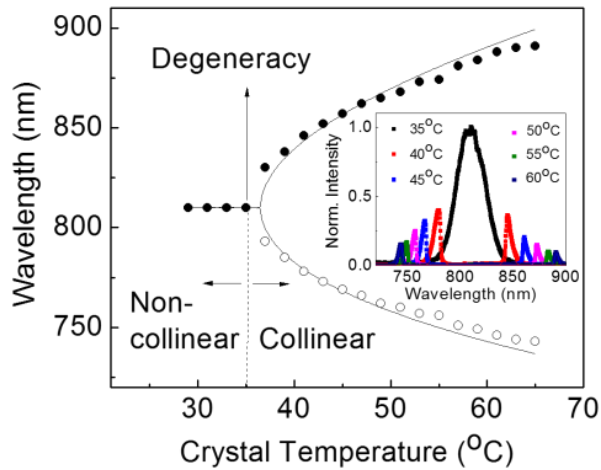

*Fig. 3. Wavelength tuning of the SPDC photons as a function of crystal temperature at collinear and non-collinear phase-matching geometries. (Inset)Emission spectra of the SPDC photons recorded at different crystal temperatures.*

After successful generation of degenerate, non-collinear SPDC photons using type-0 phase-matched PPKTP crystal, we characterized the source in terms of singles, and coincidence counts. Using the detection assembly, comprising with a collimator of 5mm in diameter and a multi-mode fiber connected to SPCM collecting photons from two diametrically opposite points of the SPDC rings formed by a convex lens of focal length, $f$=200 mm in $f$-$f$ optical system configuration, we have measured the variation in both the rate of pair production and coincidence counts with pump power at two different crystal temperatures. The results are shown in Fig. 4. As evident from Fig. 4, the coincidence counts (solid circles) and corresponding single counts (open circles) of the individual detector increases linear to the pump power at a slope of, $N_c$=24.31±0.01 x$10^4$Hz/mW and $N_1$~$N_2$=21.69±0.01x$10^5$Hz/mW respectively, at crystal temperature of 22ºC. Using the equation, $N_{Pair}$=($N_1 \times N_2$)/$N_c$[12], the number of generated photon pairs can be estimated to be as high as ~19.35±0.02MHz/mW. On the other hand, increase in crystal temperature from 22º C to 29ºC results in significant increase in the coincidence counts (solid circle) and single counts (open circle) rate at slope of $N_c$=43.37±0.06x $10^4$Hz/mW and $N_1$~$N_2$=41.20±0.02 x $10^5$Hz/mW respectively, at an estimated photon pair rate of 39.13±0.04 MHz/mW. As compared to the previous report on non-collinear degenerate SPDC source based on periodically poled LiTaO$_3$ crystal [12], our source produces ~13 times more paired photons even in the presence of comparatively loose focusing. Such increase in the photon pairs can be attributed to the use of first order grating period of the PPKTP crystal resulting three times higher effective nonlinearity, and also the use of 1.5 times longer crystal length as compared to the Ref.[12].Although, we can increase the photon pair rate further through the use of optimized pump focusing, to the best of our knowledge, this is so far the highest reported detected pairs of the bulk crystal based SPDC sources. The detection efficiencies are calculated to be ~10.7% and ~11.1% at crystal temperatures 22°C and 29°C, respectively. Since the parametric gain is almost constant near the degeneracy and the SPDC is a very feeble nonlinear process, one can in principle, expect the number of the detected pair photons to be constant with crystal temperature but increase linearly with the pump power. While the linear dependence of the coincidence (solid lines) and single (dotted lines) count rates with the pump power are maintained (see Fig. 4), we observe significant change in both the coincidence and single counts with the increase of crystal temperature. To understand such effect, we measured the coincidence count rate of the pair photons while changing the crystal temperature at a fixed pump power of 0.2 mW, with results shown by the inset of Fig. 4. As evident from the inset of Fig. 4, the coincidence counts (solid circles) increase from $0.62 \times 10^5$ Hz to $1.29 \times 10^5$ Hz with the change in crystal temperature from 22°C to 29°C, showing a quadratic dependence with the crystal temperature. Since, the coincidence counts do not have direct mathematical relation to the crystal temperature, we measured the dependence of SPDC ring radius on crystal temperature. Using a lens of focal length, $f$=200 mm in $f$-$f$ configuration, we observed the SPDC ring radius (open circles, inset of Fig. 4) to decrease linearly (dashed line) from 7.5 mm to 5.2 mm with the increase of crystal temperature from 22°C to 29°C and the inverse square of the SPDC ring radius (solid line) exactly fitting with the coincidence counts data (solid dots, inset of Fig. 4). Such effect can be understood as follows. Given that the number of SPDC photons is constant at constant pump power, and distributed over the annular area of the SPDC ring, the linear decrease in the radius of the SPDC ring with the increase in crystal temperature from 22°C to 29°C results in a quadratic increase in the number density (number of photons per unit area per sec). Therefore, the singles and coincidence counts measured using the fixed aperture sized detection system show quadratic increase with the crystal temperature. Here, we can conclude that for high brightness SPDC source one need to adjust the crystal parameters to access the optimum size of the SPDC ring. One can in principle, adjust the grating period of the PPKTP crystal to access such high brightness of the SPDC source at any favourable crystal temperature.

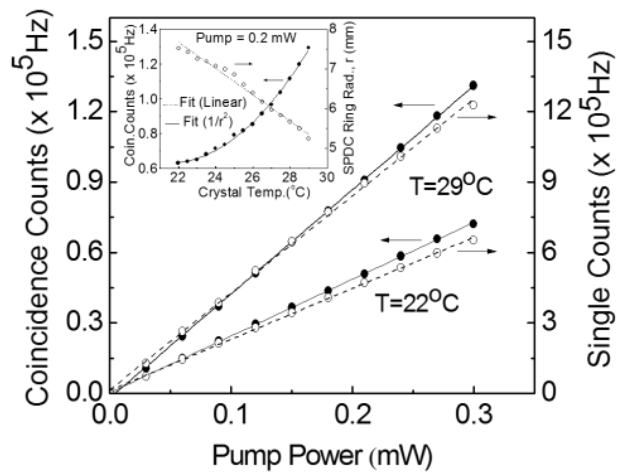

*Fig. 4. Dependence of coincidence and single count rate of SPDC photons on input pump power at two different crystal temperatures. (Inset) Variation of pair photons rate and the radius of SPDC ring as a function of crystal temperature. Lines are best fit to the experimental results.*

With the successful generation of paired photons with high brightness, we study the source for the generation of polarized entangled photons. Since in type-0 phase-matching geometry, the generated paired photons have the same state of polarization as that of the pump photons, we placed the PPKTP crystal inside the polarization Sagnac interferometer and pumped in both CW and CCW direction sat same state of polarization, vertical ($|V\rangle$). The non-collinear SPDC photons generated in both the directions are superposed using the PBS after changing the polarization state of the photons generated in one of the directions. To verify on-collinear generation of SPDC photons in both the directions, and their superposition as required for entanglement, we have recorded the intensity distribution of the SPDC photons using an EMCCD (ANDOR, DU-897U-CS0-BVAntiF) at different experimental conditions. The results are shown in Fig. 5. Adjusting the polarization state, horizontal ($|H\rangle$), and or vertical ($|V\rangle$), of the pump beam to the PBS, we pumped the PPKTP crystal in CW and or CCW directions, respectively. From Fig. 5(a) and 5 (b) showing the spatial distribution of the SPDC pairs of polarization state, $|HH\rangle$ and $|VV\rangle$ generated in CW and CCW, respectively. It is evident that the PPKTP crystal in the current experimental architecture produces indistinguishable SPDC rings of orthogonal polarization states which is further confirmed from Fig. 5 (c) and Fig. 5 (d) representing the images of the SPDC rings under misalignment and perfect alignment of the source, respectively.

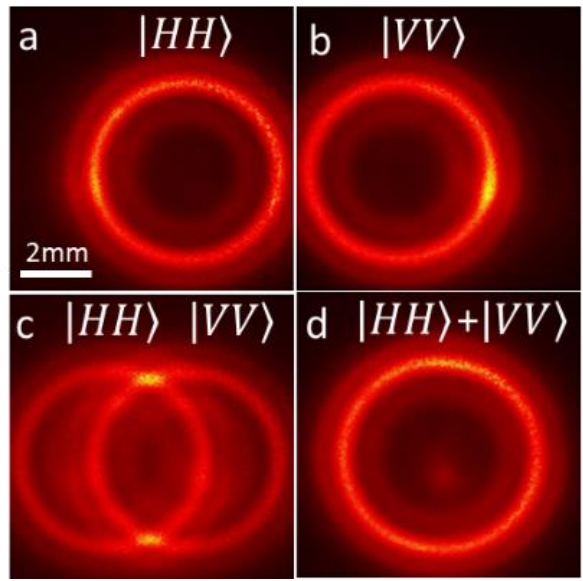

*Fig. 5. Spatial distribution of the SPDC photon pairs recorded using EMCCD. The pump power is 5mW. Images of the paired photons of polarization state, (a)$|HH\rangle$, (b) $|VV\rangle$, (c) both $|HH\rangle$ and $|VV\rangle$ with misalignment, and (d) perfectly overlapped states, $|HH\rangle$ and $|VV\rangle$.*

For quantitative analysis of the polarized entanglement of the source, we pumped the crystal with a total power of 1mW and aligned system to maintain perfect overlapping of the SPDC rings. Using a detection system comprising with an aperture of diameter ~0.5mm, an interference filter of bandwidth~2nm, analyzer, A, multimode fiber, MMF, and SPCM we selected photons from two diametrically opposite points of the SPDC ring and analyzed the polarization correlation function in two mutually unbiased bases, horizontal, H, and diagonal, D, at two crystal temperatures. The Fig.6, shows the quantum interference of the polarized entangled photons for H(closed and open circles) and D(closed and open triangles) projections for the crystal temperature $23^\circ C$ (open circles and triangles) and $25^0C$ (closed circles and triangles). The solid and dotted lines are best fit to the experimental data. The interference visibility in H/D projections are estimated to be 94.8±0.5%/71.3±0.3% and 96.9±0.4%/75.3±0.3% for the crystal temperature of $25^0C$ and $23^0C$, respectively. The measured visibilities under both basis are higher than 71%, large enough to violate Bell's inequality. The errors are estimated based on Poisson statistics. The reduction in the coincidence count can be attributed to the small collection aperture size and narrow spectral filters. Using the interference graph, we can identify the generated state to be, $\psi = 1/\sqrt{2}(|HH\rangle + e^{-i\phi}|VV\rangle)$. The phase, $\phi$, can be controlled by proper positioning of the crystal inside the interferometry and for the current setup we optimized to π. The Bells parameter, S, at crystal temperature 25$^0$C and 23$^0$C is calculated to be, S= 2.29±0.03 and S=2.39±0.04, respectively, clearly violating the Bell's inequality by nearly 10 standard deviations. Additionally, we assess the degree of entanglement through the quantum state tomography, resulting a Bell state fidelity of 0.96. The lower visibility in diagonal projection can be improved further by using proper phase compensation schemes. Similar to the report[12], we also observe a drastic reduction of entanglement visibility with the increase of pump power beyond 2 mW might be due to multi-pairs generation at increased source rate.

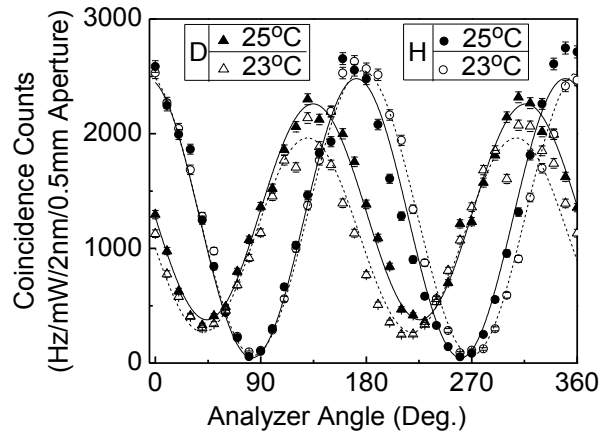

*Fig. 6 Quantum interference of the entangled photons at horizontal, H (closed and open circles) and diagonal, D (closed and open triangles) projection bases for crystal temperature 25$^o$C (closed circle and triangles) and 23$^o$C (open circle and triangles). The lines are the best fit to the experimental data.*

In conclusion, we have demonstrated a simple, and compact source of entangled photons at high brightness. Use of non-collinear generation of degenerate SPDC photons through type-0 phase-matching of a single PPKTP crystal at room temperature in a Sagnac interferometer, makes the source robust, simple and importantly low cost. The source produces a detected pair rate of 39.13±0.04MHz/mW even in the presence of loose focusing, so far the highest reported detected pair rate from bulk crystal based SPDC sources. The polarization correlation study and quantum tomography measurement revels that the source produces entangled state violating the Bell's inequality by nearly 10 standard deviations and a Bell state fidelity of 0.96. Such high brightness entangled photon source in compact and rugged architecture is ideal for the many present and future experiments in the field of quantum optics especially in quantum communications.